\documentclass[numberedappendix,12pt,preprint]{emulateapj}
\slugcomment{Accepted by the Astrophysical Journal, September 25, 2015} 

\usepackage[]{amsmath}

\shorttitle{The galaxy LF before reionization}
\shortauthors{Mason, Trenti \& Treu}

\usepackage{color}			   
\definecolor{midgray}{gray}{0.4}		
\definecolor{orange}{rgb}{1,0.5,0}  

\usepackage[colorlinks=true,citecolor=midgray,linkcolor=midgray]{hyperref}    

\newcommand{\simgt}{\,\rlap{\lower 3.5 pt \hbox{$\mathchar \sim$}} \raise
1pt \hbox {$>$}\,}
\newcommand{\simlt}{\,\rlap{\lower 3.5 pt \hbox{$\mathchar \sim$}} \raise
1pt \hbox {$<$}\,}

\newcommand{\nh}{\langle n_{\textrm H}\rangle}
\newcommand{\nion}{\dot{n}_{\textrm ion}}
\newcommand{\trec}{t_{\textrm rec}}
\newcommand{\Ob}{\Omega_{\textrm b}}
\newcommand{\Om}{\Omega_{\textrm m}}
\newcommand{\OL}{\Omega_{\Lambda}}

\newcommand{\lya}{Ly$\alpha$}

\newcommand{\HST}{\textit{HST}}
\newcommand{\JWST}{\textit{JWST}}
\newcommand{\WFIRST}{\textit{WFIRST}}

\newcommand{\Eq}[1]{Equation~(\ref{#1})}

\newcommand{\eps}{$\varepsilon(M_h)$}

\newcommand{\BE}{\begin{equation}}
\newcommand{\EE}{\end{equation}}
\newcommand{\BEA}{\begin{eqnarray}}
\newcommand{\EEA}{\end{eqnarray}}

\begin{document}

\title{The Galaxy UV Luminosity Function Before the Epoch of Reionization}

\author{
Charlotte A. Mason$^{1,2}$,
Michele Trenti$^{3}$, and 
Tommaso Treu$^{1,2}$
}
\affil{$^{1}$ Department of Physics, University of California, Santa Barbara, CA, 93106-9530, USA}
\affil{$^{2}$ Department of Physics and Astronomy, UCLA, Los Angeles, CA, 90095-1547, USA}
\affil{$^{3}$ School of Physics, University of Melbourne, Parkville, Victoria, Australia}
\email{cmason@physics.ucsb.edu}

\begin{abstract}

  We present a model for the evolution of the galaxy ultraviolet (UV) luminosity function (LF) across cosmic time where star formation is linked to the assembly of dark matter halos under the assumption of a mass dependent, but redshift independent, efficiency. We introduce a new self-consistent treatment of the halo star formation history, which allows us to make predictions at $z>10$ (lookback time $\simlt500$ Myr), when growth is rapid. With a calibration at a single redshift to set the stellar-to-halo mass ratio, and no further degrees of freedom, our model captures the evolution of the UV LF over all available observations ($0\simlt z\simlt10$). The significant drop in luminosity density of currently detectable galaxies beyond $z\sim8$ is explained by a shift of star formation toward less massive, fainter galaxies. Assuming that star formation proceeds down to atomic cooling halos, we derive a reionization optical depth $\tau = 0.056^{+0.007}_{-0.010}$, fully consistent with the latest Planck measurement, implying that the universe is fully reionized at $z=7.84^{+0.65}_{-0.98}$. In addition, our model naturally produces smoothly rising star formation histories for galaxies with $L\simlt L_*$ in agreement with observations and hydrodynamical simulations.  Before the epoch of reionization at $z>10$ we predict the LF to remain well-described by a Schechter function, but with an increasingly steep faint-end slope ($\alpha\sim-3.5$ at $z\sim16$). Finally, we construct forecasts for surveys with \JWST~and \WFIRST and predict that galaxies out to $z\sim14$ will be observed. Galaxies at $z>15$ will likely be accessible to \JWST~and \WFIRST~only through the assistance of strong lensing magnification. \end{abstract}


\section{Introduction}
\label{sec:intro}

The rest-frame ultraviolet (UV) galaxy luminosity function (LF) and its evolution with redshift are crucial tracers of galaxy properties over cosmic time. In particular, UV light can be used efficiently to measure the star formation rate (SFR), because photons at rest-frame wavelengths around 1500 \AA~are primarily produced by young, massive, and short-lived stars. Current observations characterize the UV LF over the large majority of the history of the universe, ranging from studies in the local universe from \emph{Galaxy Evolution Explorer} data (e.g.,~\citealt{Burgarella2006} to \emph{Hubble Space Telescope Wide Field Camera 3} (\HST/WFC3) observations which now reach redshift $z\sim8-10$, i.e. lookback times greater than 13 Gyr (e.g.~\citealt{Bouwens2015a}). Transformational results on the rest-frame UV light from the epoch of reionization, when the universe was less than 0.8 Gyr old, have been made possible by large and dedicated community efforts that have identified a sample of more than $1000$ galaxy candidates at $z>7$, spanning a large range in luminosities. This progress is thanks to a variety of surveys, including ultradeep observations in blank fields with the HUDF09 and HUDF12 campaigns~\citep{Bouwens2011,Dunlop2013,Illingworth2013}, the use of cluster-scale lensing to probe intrinsically faint objects~\citep[e.g.,][]{Yue2014a,Atek2015}, the large area, panchromatic CANDELS survey~\citep{Grogin2011,Koekemoer2011}, and wide-field random-pointing surveys to identify more luminous but rarer objects at the bright end of the LF~\citep[e.g.,][]{Trenti2011,Bradley2012,Schmidt:2014p34189}.

The picture emerging from these observations is that the number density of galaxies decreases with increasing redshift, while the LF remains consistent with a~\citet{Schechter1976} form, $\Phi(L) = \Phi^* (L/L^*)^\alpha \exp{(-L/L^*)}/L^*$ out to $z\sim 8$, albeit the uncertainty and co-variance in the best fitting parameters grows at high redshift~\citep{Schmidt:2014p34189,Bouwens2015a}. However, other studies find either an equally good fit with a double-power law at $z\sim 8$ \citep{Finkelstein2015}, or a preference for the double power law at $z\sim 7$ (\citealt{Bowler2014a,Bowler2015}; but note that the $z\sim7$ investigations are based on ground, rather than space observations). Irrespective of the form of the LF at the bright-end, whose evolution might be linked to changing astrophysical conditions of high-redshift galaxies, such as reduced feedback at early times \citep{Somerville2008}, it is clear that overall, the observed population of galaxies ($M_{AB}\simlt -17$) at $z>7$ does not produce sufficient photons to ionize the universe. This is consistent both with theoretical modeling~\citep{Trenti2010,Raivevic2011,Robertson2015} and indirect probes, such as the lack of detections of gamma ray burst host galaxies~\citep{trenti2012_GRB,Tanvir2012}, which suggest that even the deepest \HST~observations are seeing only the tip of the iceberg of the population of star-forming galaxies. Therefore, it is likely that a very faint population of unseen dwarf galaxies at $z\simgt 6$ is the main contributor to the UV luminosity density and in turn to the ionizing photon budget (e.g., see~\citealt{Alvarez2012}). Indirect observational support for extrapolation of the UV LF is provided by the detection of ultrafaint galaxies behind gravitational lenses at $z\sim 2$~\citep{Alavi2014}. However, see ~\citet{Giallongo2015} and ~\citet{Madau2015} for a discussion of the potential contribution of active galactic nuclei to reionization.

\HST~is both wavelength and aperture limited to observations at $z\simlt11$, but future progress to characterize the properties of the UV LF at high redshift will be boosted by the upcoming James Webb Space Telescope ~\citep[\JWST;][]{JWST_SSR} and Wide-Field Infrared Survey Telescope ~\citep[\WFIRST;][]{Spergel2015}. These observatories are expected to extend the frontier of galaxy detection to before the epoch of reionization at $z>10$, when the first generation of galaxies were being assembled and hydrogen in the universe was predominantly neutral.

Motivated by this upcoming improvement in discovery capabilities of high redshift galaxies, we aim here at predicting the UV LFs at $z>10$. For this we use a simple, yet successful, semi-analytic framework introduced previously by~\citet{Trenti2010,Trenti2015,Tacchella2013} which assumes that the main driver of the evolution of the galaxy UV LF is the growth and hierarchical assembly of dark matter (DM) halos. Whilst this simple approach does not have the power of numerical hydrodyamic simulations ~\citep[e.g.,][]{Genel2014,Furlong2015} to make predictions about galaxy properties such as morphology, and the role played by satellite galaxies, it provides a simple, robust and empirically calibrated method to make predictions for the evolution of global galaxy properties. In particular, the simplicity of our approach avoids the degeneracies of large multi-parameter numerical simulations, and allows us to calibrate at one reference redshift the complex physics that regulates the conversion of baryons into stars, and then to focus on investigating how DM halo assembly drives redshift evolution under the assumption that star formation efficiency is redshift independent at fixed halo mass. 

The key assumptions of the framework are that, for star forming galaxies, the stellar mass content of a DM halo depends on the halo mass but not on redshift, and that the stellar mass built up has a characteristic timescale given by the halo assembly time. The mass dependent efficiency of converting halo mass into stellar mass is calibrated empirically at one redshift, where good observational constraints are available (e.g. $z=4-5$), and then applied to predict the redshift evolution of the UV LF by combining evolution in the DM halo mass function (HMF) and halo assembly time with stellar population synthesis modeling. Since halos assembled more rapidly in the past, the SFR was higher at high redshift, explaining the general brightening of the luminosity versus halo mass relation found by abundance matching studies (e.g.~\citealt{Cooray2005}).

This simple strategy has been remarkably successful in describing the UV LF evolution with redshift~\citep{Tacchella2013}, with results similar to those obtained by other studies based on matching DM halos to galaxy luminosity (e.g.~\citealt{Lacey2011,Munoz2012,Jaacks2013,Behroozi2014,Dayal2014,Mashian2015}). However, the earlier implementations had two limitations. First, the model took into account only the star formation happening in the second half of the halo assembly history: the time the halo took to grow from $M_h/2$ to $M_h$. In addition, the calibration of the star formation efficiency was not guaranteed to be internally self-consistent with the halo assembly history of a DM halo over redshift. Given that the UV luminosity of a galaxy is only weakly sensitive to the star formation history of stellar populations with ages greater than a few hundred Myr~\citep{Madau&Dickinson2014}, these limitations had very little impact at $z\simlt 8$. However the situation is potentially different at early times, when the assembly time goes below 100 Myr. Thus, the previous version of the model could not be trusted to formulate predictions at $z\geq 10$.

In this paper, we develop and present an improved and self-consistent model that describes the full star formation history of a DM halo, and we verify that (1) it continues to describe well the UV LF at $z\simlt8$; (2) it successfully reproduces the latest observations at $z\simgt 8$, providing an important validation of the approach introduced before such observations were available. In addition, we make detailed predictions for future surveys at $z\geq 10$ with the James Webb Space Telescope (\JWST) and the Wide-Field Infrared Survey Telescope (\WFIRST), and we discuss the implications of our results in the context of the ionizing photon budget.

This paper is organized as follows: in Section~\ref{sec:model} we introduce our new model and its calibration; Section~\ref{sec:results} describes our model results, and our predictions for future surveys; and we summarize and conclude in Section~\ref{sec:conc}. All magnitudes are AB magnitudes and we use the ~\citet{Planck2015cosmo} cosmology, with $\Om=0.315$, $\OL=0.685$, $\Ob=0.0490$, $h=0.6731$, $\sigma_8=0.829$ and $n_s=0.9655$.

\section{Model Description}
\label{sec:model}

Our model considers the growth of DM halos to be the most important driving force in the growth of galaxies, and aims at predicting the evolution of the UV luminosity function with a minimal number of assumptions. We thus assume that the SFR is proportional to: (1) the halo mass, through a mass-dependent but redshift-independent efficiency \eps, which is the ratio of the stellar mass formed during the halo assembly time $t_a(M_h,z)$ to the final halo mass; and (2) to the inverse of the halo assembly time. The halo assembly time, $t_a(M_h,z)$, of a halo of mass $M_h$ observed at redshift $z$ is the lookback time at which the progenitor halo had mass $M_h/2$~\citep{Lacey1993}, and decreases at higher redshift. This implies that DM halos at fixed mass host star forming galaxies with stellar mass which is independent of redshift, but with stellar populations that are younger at higher redshift. 

Since the HMFs and assembly times are fully defined by the cosmological model parameters, these assumptions allow us to calibrate our model at one redshift to derive \eps, and then construct predictions for the galaxy luminosity function at all other redshifts from the DM HMF and the halo assembly time.

\subsection{Star formation prescription}
\label{sec:model_SF}

The UV luminosity of a galaxy most strongly depends on its youngest stars, while stellar populations older than a few hundred Myr contribute little. However, at $z \simlt 8$, the halo assembly time is less than 100 Myr, so to predict the UV luminosity at high redshift it is necessary to consider periods of star formation before the halo assembly time.

To include multiple epochs of star formation, as a halo grows in DM mass, we define the star formation history of a halo as a linear combination of constant bursts normalized by the length of each burst. Thus we define the SFR for a halo with mass $M_h$  in each epoch between times $t_i$ and $t_{i+1}$ as:
\BE  \label{eqn:model_sfr}
	SFR(t_i,t_{i+1}, M_h) = M_h \times \frac{\varepsilon(M_h/2^i)}{2^i (t_{i+1}-t_i)} 	
\EE
where we define $t_0$ as the lookback time for a halo observed at redshift $z_\textrm{obs}$ and $t_{i>0}=t_a(M_h/2^{i-1}, z_{i-1})$, where $t_a$ $(z_a)$ is the halo assembly time (redshift). We similarly define $z_0 = z_\textrm{obs}$ and $z_{i>0} = z_a(M_h/2^{i-1}, z_{i-1})$. We calculate the halo assembly time as defined by ~\citet{Lacey1993} in the extended Press-Schechter formalism~\citep{Bond1991} using an ellipsoidal collapse model~\citep{Sheth2001,Giocoli2007}. We use the median of the probability distribution of assembly times for each halo. While this assumption does not take into account variations in the luminosity of individual galaxies, there is a minimal effect on the global LF from neglecting scatter in halo assembly times, as demonstrated by ~\citet{Tacchella2013}. 

\begin{figure}[!t] 
\includegraphics[width=0.46\textwidth]{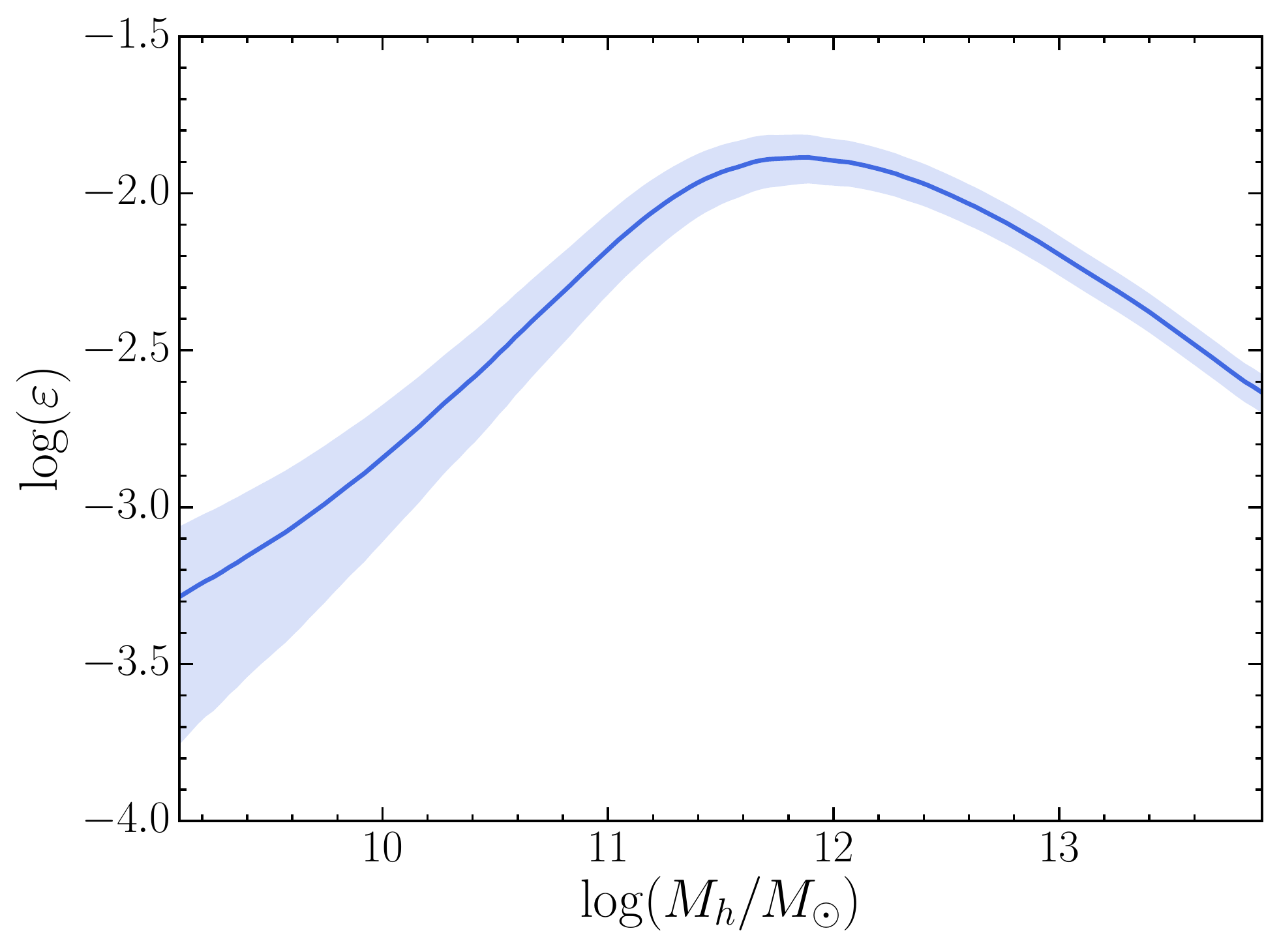}
\vspace{0.3cm}
\caption{The efficiency of star formation, the ratio of stellar mass formed during the halo assembly time to halo mass, see \Eq{eqn:model_mstell}, derived at the calibration redshift $z\sim5$, as described in Section~\ref{sec:model_calib}. The shaded region shows $1\sigma$ confidence range.}
\label{fig:results_eps}
\vspace{0.5cm}

\end{figure}

We define the redshift-independent efficiency of star formation, \eps, as the ratio of the stellar mass formed during the halo assembly time to the final halo mass. Thus, to make predictions, we only need to calibrate \eps~at one redshift (see Section~\ref{sec:model_calib} and Figure~\ref{fig:results_eps})  and can use the derived \eps~for all further predictions.

\begin{figure}[!t] 
\includegraphics[width=0.49\textwidth]{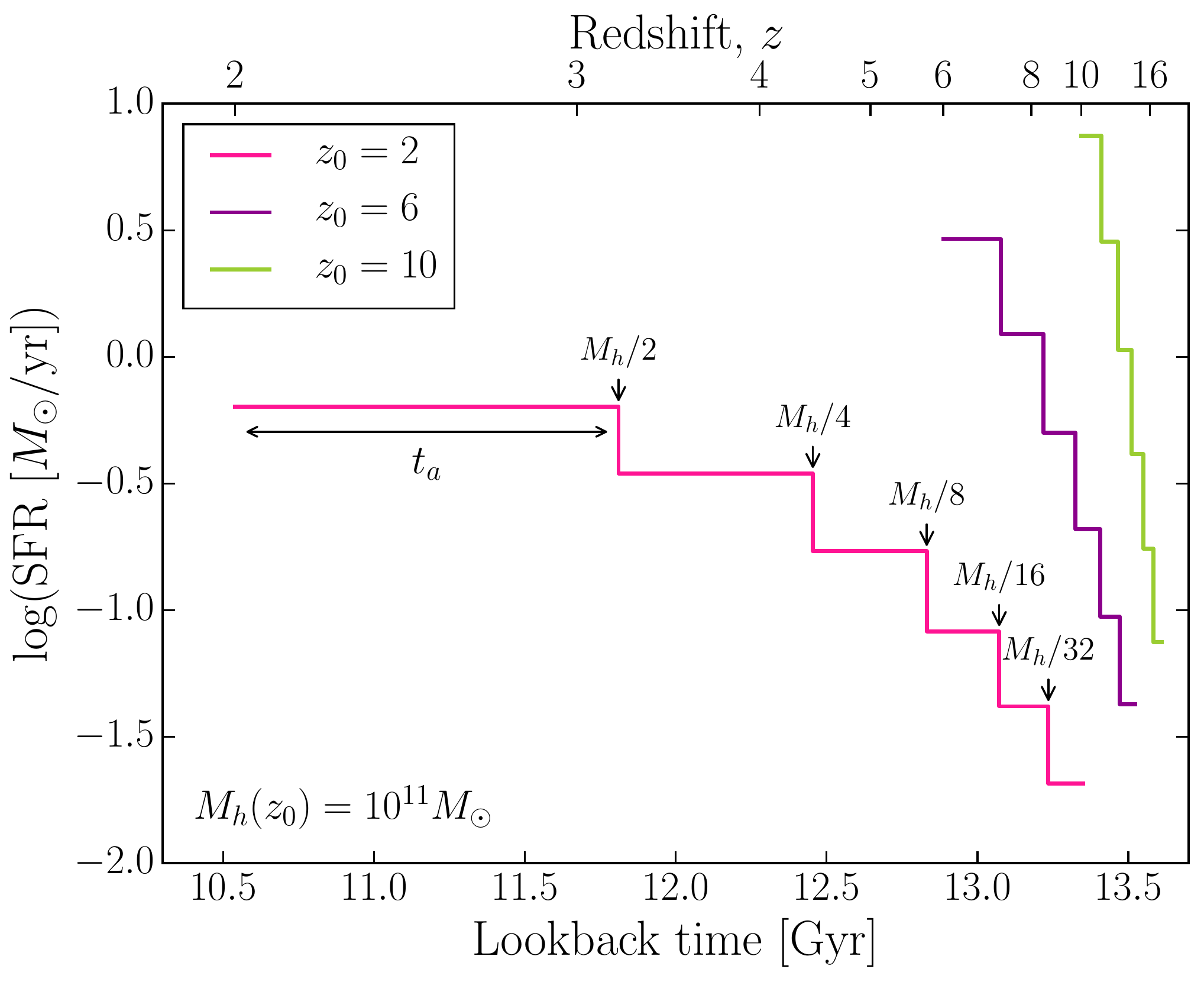}
\\
\caption{The star formation history, as described by \Eq{eqn:model_sfr} in our model, of a halo of fixed mass $10^{11}~M_\odot$, if observed at $z_0=2$, $z_0=6$ or $z_0=10$. We label the halo assembly time for the $z_0=2$ halo and the mass of the halo progenitor at the start of each constant star formation epoch.}
\vspace{0.5cm}
\label{fig:model_sfh}
\end{figure}

Figure~\ref{fig:model_sfh} shows the star formation history of halos of fixed final mass $M_h = 10^{11}~M_\odot$ observed at $z_\textrm{0}=2, \, 6$ and $10$, calculated using the SFR in \Eq{eqn:model_sfr}. As \eps~is redshift independent, these halos will also have identical stellar masses at their observed redshifts. The SFR shown in Figure~\ref{fig:model_sfh} increases in each epoch as the halo grows from $M_h/64$ to $M_h$, because \eps~decreases with decreasing halo mass (for $M_h\simlt10^{12}~M_{\odot}$) more rapidly than the shortening of the halo assembly times as the lookback time grows. This behavior of our model is fully consistent with strong evidence of rising star formation histories with redshift from both numerical simulations and observations~\citep{Finlator2011,Papovich2011,Jaacks2012,Behroozi2013b,Lee2014}. Thus the greatest contribution to the stellar mass is during the halo assembly time as the halo grows from $M_h/2$ to $M_h$. This figure also illustrates how the short halo assembly times at high redshift require a considerably higher SFR to form the same final stellar mass.

We include the contribution from star formation in successively smaller halo progenitors by summing the terms from \Eq{eqn:model_sfr}. The sum is truncated when the progenitor halo mass is below the cooling threshold, i.e. $\varepsilon(M_h \simlt 10^8M_{\odot})=0$.

Thus we can derive the stellar mass as:
\BE  \label{eqn:model_mstell}
M_\star(M_h) = M_h \times \sum_{i=0}^{i=\infty} \frac{\varepsilon(M_h/2^i)}{2^i}
\EE
which is redshift independent.

To compute the UV luminosity of a halo we populate every halo with a galaxy with a stellar population based on the simple stellar population (SSP) models of ~\cite{Bruzual2003}. We assume a Salpeter initial mass function (IMF) between $0.1M_\odot$ and $100M_\odot$, as low mass stars do not contribute much to UV luminosity, and constant stellar metallicity $Z=0.02Z_\odot$. We neglect redshift evolution in metallicity as the UV luminosity does not depend strongly on metallicity under the assumption that current and future HST/JWST surveys detect galaxies living in relatively massive DM halos ($M_h\simgt 10^{9}~\mathrm{M_{\odot}}$) where multiple generations of early star formation began enriching the gas at $z\simgt 20-40$ \citep{Bromm2009,Trenti2009,Smith2015}, so that by $z\simlt 16$ the metallicity has risen to $Z\simgt 0.01~Z_{\odot}$. We define $\ell_\textsc{bc}(t)$ as the luminosity at 1500 \AA~of an SSP of mass $1 M_\odot$ and age $t$. The total UV luminosity of a galaxy observed at redshift $z$ is obtained by integrating over the SFR (\Eq{eqn:model_sfr}) and SSP luminosity in each epoch of star formation:
\BE  \label{eqn:model_Lpred}
L(M_h,z) = \sum_{i=0}^{i=\infty} \int_{t_i}^{t_{i+1}} SFR(t_i,t_{i+1},M_h) \ell_\textsc{bc}(t) dt
\EE
Where $t_i$ are the halo assembly times for the half-mass progenitors defined above.


\subsection{Dust extinction}
\label{sec:model_dust}

The observed UV luminosity is significantly attenuation by dust extinction, particularly at $z\simlt4$. Thus, we include dust extinction in our model, following closely the procedure adopted in observations of Lyman-break galaxies. We assume a spectrum modeled as $f_\lambda \sim \lambda^\beta$, and extinction $A_\textsc{uv} = 4.43 + 1.99\beta$~\citep{Meurer1999}. Following ~\citet{Trenti2015} and~\citet{Tacchella2013} we model the observations of $\beta$ by~\citet{Bouwens2014} as:
\BEA  \label{eqn:model_dust}
&&\langle \beta(z,M_\textsc{uv}) \rangle = \\
	 &&\left\{ \begin{array}{l l}
   \left(\beta_{M_0}(z) - c\right) \exp \left[-\frac{\frac{d\beta}{dM_0}(z)[M_\textsc{uv}-M_0]}{\beta_{M_0}(z) - c}\right] + c & \quad M_\textsc{uv} \geq M_0 \\
   \frac{d\beta}{dM_0}(z)[M_\textsc{uv} - M_0] + \beta_{M_0}(z) & \quad M_\textsc{uv} < M_0
   \end{array} \right. \nonumber
\EEA
where $c=-2.33$, $M_0 = -19.5$, and the values of $\beta_{M_0}$ and $d\beta/dM_0$ are taken from Table 3 in~\citet{Bouwens2014} and linearly extrapolated to lower and higher redshifts. Using this linear plus exponential model for $\beta$ we fit the~\citet{Bouwens2014} observations well - which show evidence for a curved relation between $\beta$ and $M_\textsc{uv}$. The exponential fit at faint magnitudes avoids unphysical negative dust corrections.

We assume a Gaussian distribution for $\beta$ at each $M_\textsc{uv}$ value (with dispersion $\sigma_\beta = 0.34$), giving the average extinction $\langle A_\textsc{uv} \rangle = 4.43 + 0.79 \ln(10)\sigma_\beta^2 + 1.99\langle\beta\rangle$. We use this average extinction to calculate observed UV fluxes in our model.

\subsection{Calibration}
\label{sec:model_calib}

We calibrate the model by finding \eps~such that $L(M_h, z_c) = L^\textrm{obs}(M_h, z_c)$, where $z_c$ is the calibration redshift. To find $L^\textrm{obs}(M_h, z_c)$ we derive an empirical relation between observed luminosity and halo mass by performing abundance matching~\citep{Mo1999} between the HMF, $n(M_h,z)$, and observed LF, $\Phi(L,z)$, assuming every halo hosts one galaxy:
\BE  \label{eqn:model_calib_match}
	\int_{M_h}^\infty n(M_h',z) \, dM_h' = \int_{L^\textrm{obs}}^\infty \Phi(L',z) \, dL'
\EE
where we use the~\citet{Sheth1999} HMF and calibrate at $z\sim5$ using the~\citet{Bouwens2015a} LF. We calibrate over the halo mass range $10^7 - 10^{14} M_\odot$, and therefore extrapolate the LF over $M_\textsc{uv} = -7.5$ to $M_\textsc{uv} = -24.5$ to perform the abundance matching.

We can then calculate the UV luminosity as a function of halo mass at any redshift using \Eq{eqn:model_Lpred}. The largest contribution to the uncertainty in the predicted luminosity is from the uncertainty in the calibration LF; we plot $1\sigma$ confidence bounds from this uncertainty in the plots which follow.

\section{Results}
\label{sec:results}

\subsection{Stellar masses and ages}
\label{sec:results_stell}

Figure~\ref{fig:results_eps} shows the efficiency of star formation as a function of halo mass, \eps, derived at the calibration redshift $z\sim5$. Halos with mass $M_h \sim 10^{11} - 10^{12} ~M_\odot$ have the highest star formation efficiencies, which is consistent with the consensus picture that low-mass halos have lower efficiencies because of supernova feedback, while star formation in higher mass halos is affected by strong negative AGN feedback. Our derived \eps~is consistent, to first approximation, with the results obtained by~\citet{Behroozi2013ApJ} who calculated the ratio of stellar to halo mass at different redshifts ($z\leq 4$) by applying abundance matching techniques between the stellar mass function and HMF. We note that \eps~is redshift-independent in our model, but that the evolution of the HMF and halo assembly times allows to us predict the UV luminosity using \Eq{eqn:model_Lpred}, via our rising star formation history (see Figure~\ref{fig:model_sfh}). Thus we find that with our simple redshift-independent \eps~the star formation at high redshift proceeds much more rapidly than at low redshift.

\begin{figure}[!t] 
\includegraphics[width=0.47\textwidth]{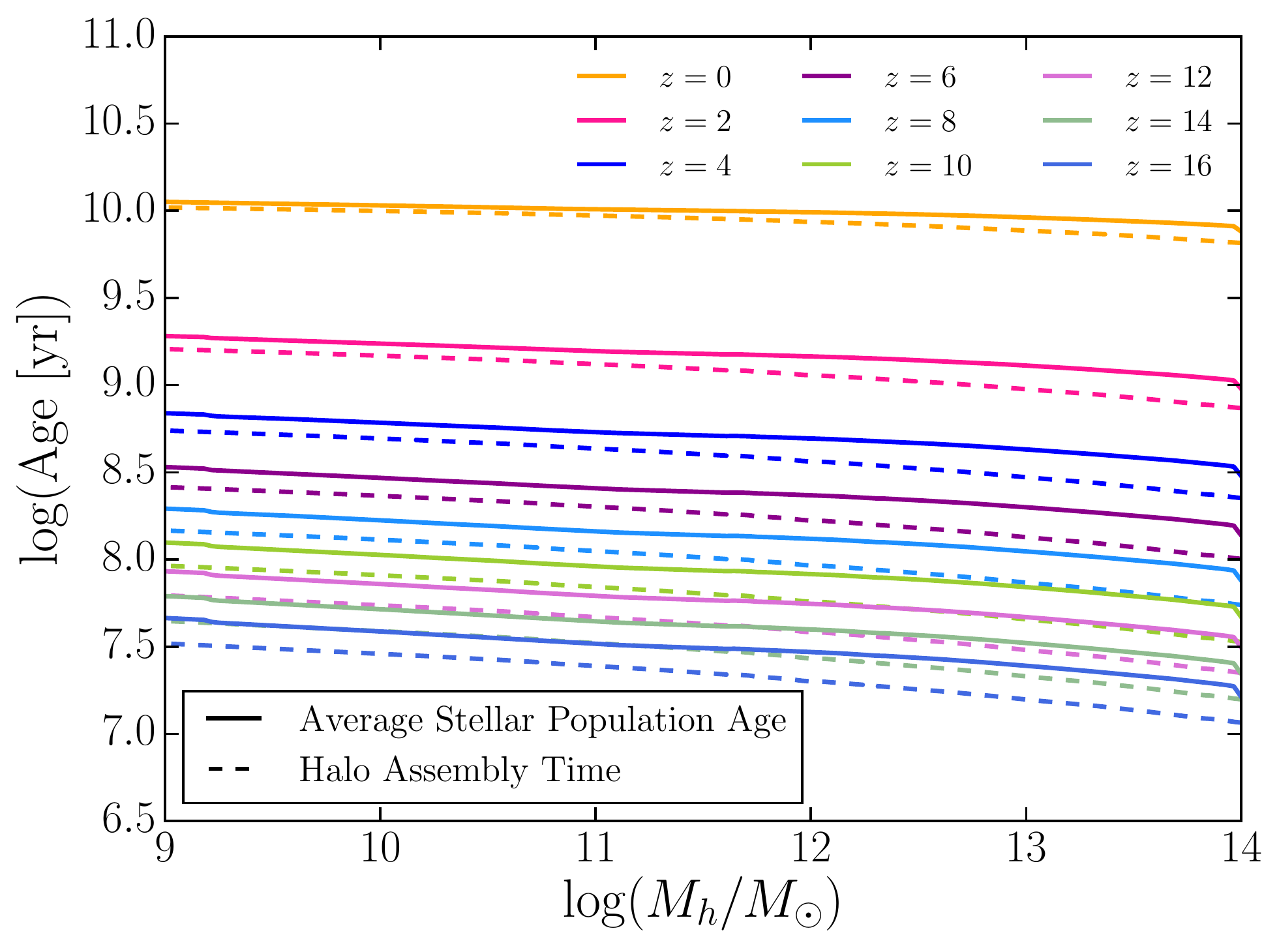}
\caption{Average stellar population age in our model and halo assembly time~\citep{Lacey1993} as a function of halo mass and observed redshift.}
\label{fig:results_ages}
\end{figure}

With Equation~\ref{eqn:model_sfr}, we can derive the model predictions for average galaxy ages as a function of halo mass and redshift, which are shown in Figure~\ref{fig:results_ages}. Galaxy ages are challenging to constrain observationally, due to the degeneracy between dust extinction, age and metallicity in spectral energy distribution (SED) fitting, which are especially severe at $z\simgt 4$ where \HST~covers only the rest-frame UV wavelengths. Our modeling results are consistent with the picture emerging from multiple recent studies that combine \HST~and Spitzer/IRAC data and find that the majority of $z\simgt4$ galaxies to have old stellar populations ($>100$ Myr) and relatively low specific SFR ($\sim 100 \,M_\odot$ yr$^{-1}$)~(\citealt{Oesch2013b,Straatman2014}; see also \citealt{Gonzalez2011} for earlier studies reaching the same conclusion but without accounting for nebular emission lines). However, other independent studies based mostly on overlapping datasets reached the different conclusion that $z\simgt 4$ galaxies have a high chance of being young systems (ages $<50$ Myr) with high specific SFRs \citep{deBarros2014,Finkelstein2015b}, suggesting that improved observational constraints are needed to evaluate the fidelity of stellar ages predicted by our model.

\begin{figure}[!t] 
\includegraphics[width=0.47\textwidth]{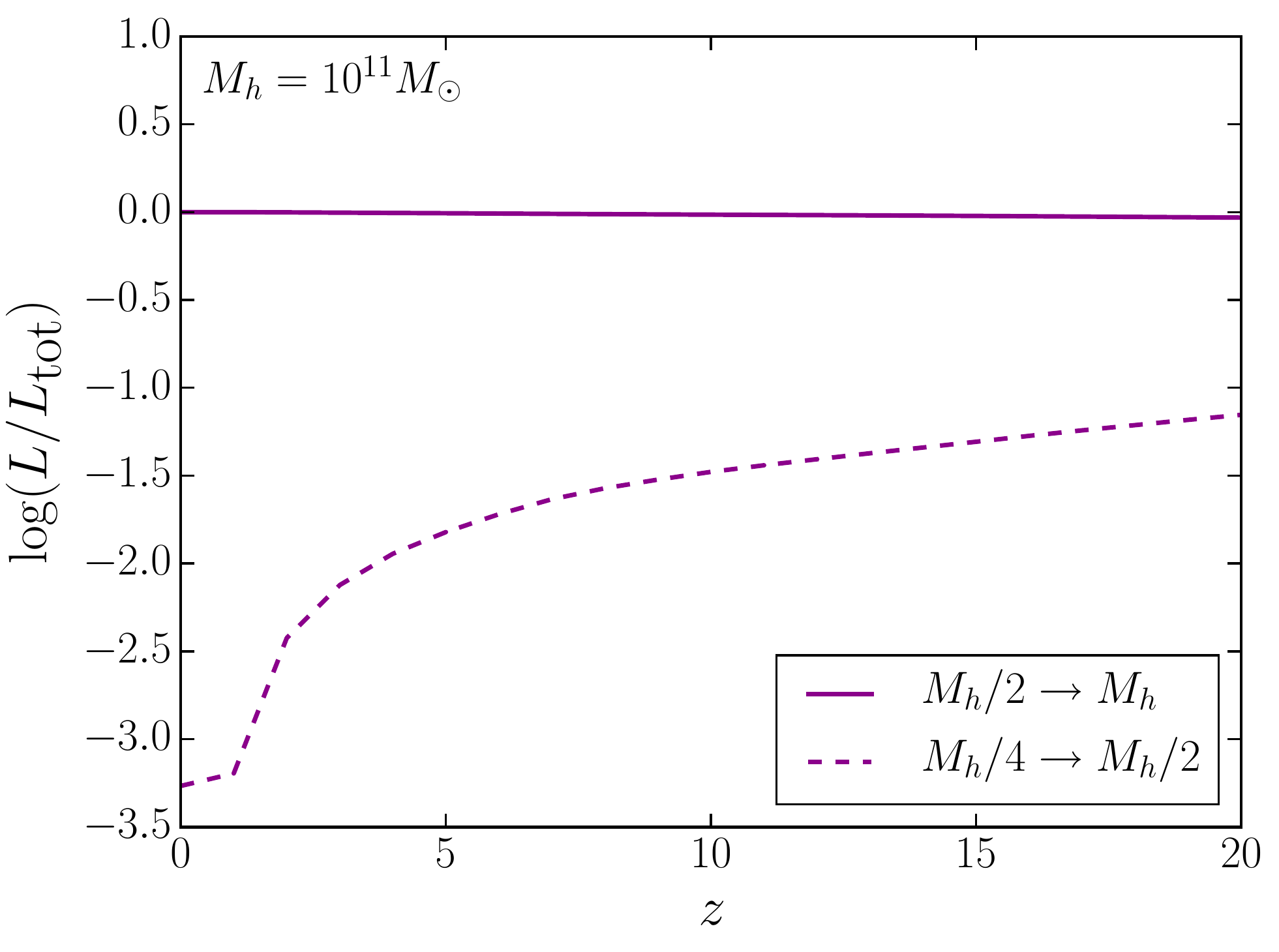}
\caption{Contribution to predicted UV luminosity from the 0th and 1st order terms in \Eq{eqn:model_Lpred} as a function of observed redshift for fixed halo mass, $M_h = 10^{11}M_\odot$. The contribution from star formation during the halo assembly time (solid) dominates, with the contribution from the earlier star formation epoch (dashed) increasing with redshift. The SFR is constant in both epochs.}
\label{fig:results_lum_z}
\end{figure}

The contributions to the total UV luminosity at fixed halo mass from two epochs of star formation as a function of observed redshift are shown in Figure~\ref{fig:results_lum_z}. We find that the earlier epoch of star formation, as the halo grows from $M_h/4$ to $M_h/2$ adds a negligible contribution to the total UV luminosity, suggesting that recent star formation is the most important contribution to the UV luminosity even at $z>10$ when the assembly time is short. This is consistent with recent clustering studies~\citep{Barone-Nugent2014} which found a high duty cycle for galaxies at $z\geq6.5$, possibly due to their bright, young stellar populations, and with the prediction of smoothly rising SFRs from numerical simulations~\citep{Finlator2011} and observations~\citep{Papovich2011,Behroozi2013b,Lee2014}. Thus, we only consider the contribution from the first two terms in \Eq{eqn:model_Lpred} in predicting the UV luminosity.

\begin{figure}[!t] 
\includegraphics[width=0.47\textwidth]{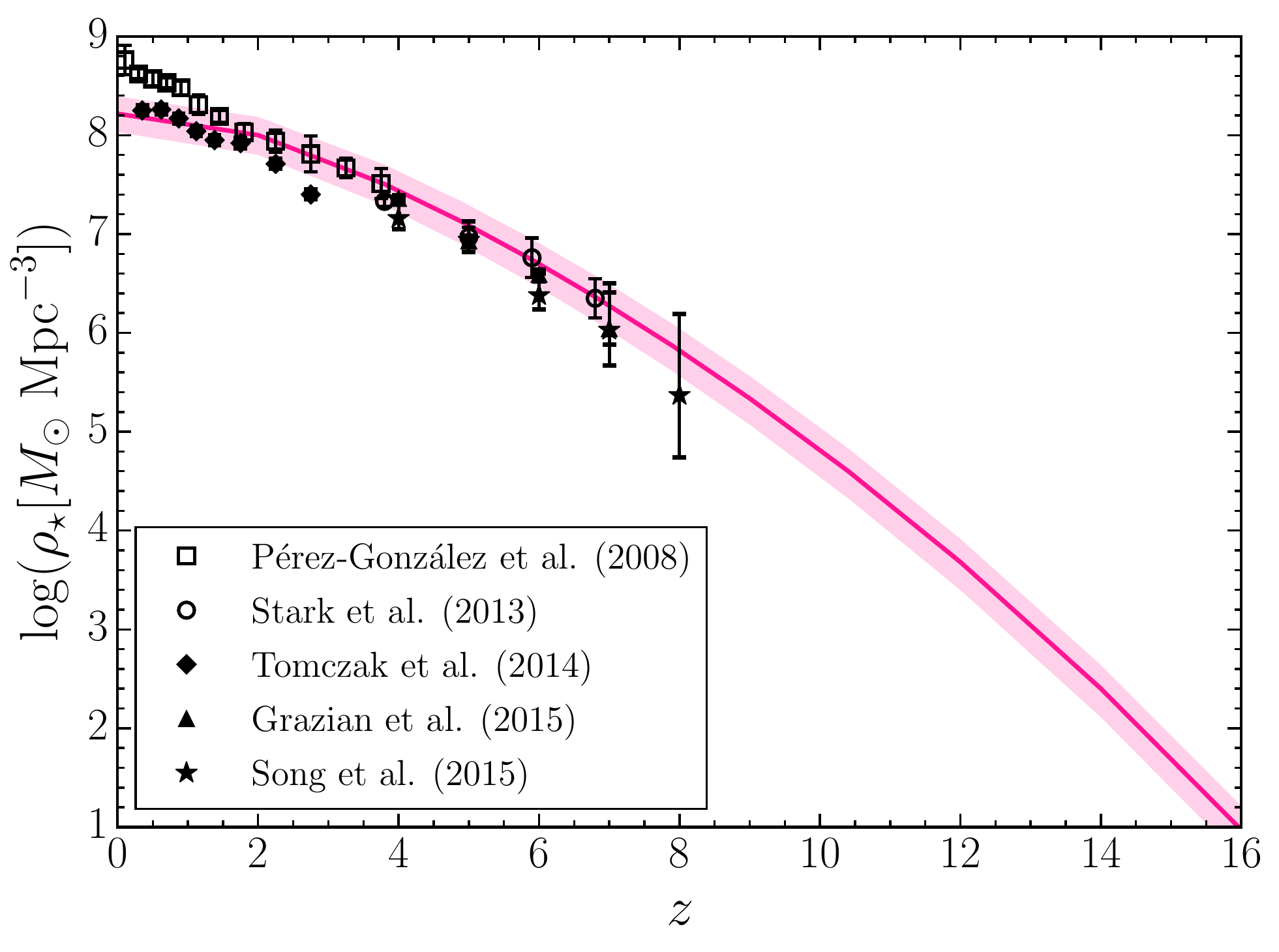}
\caption{Stellar mass density ($\rho_\star$) obtained by integrating the stellar mass function derived from our model using \Eq{eqn:model_mstell} to a stellar mass limit of $M_\star > 10^8 M_\odot$. We plot observations from~\citet{PerezGonzalez2008,Stark2013,Tomczak2014,Grazian2015,Song2015}. Shaded regions show the $1\sigma$ confidence range.}
\label{fig:results_smd}
\end{figure}

The stellar mass density as a function of redshift, obtained by integrating our model stellar mass functions (using \Eq{eqn:model_mstell} to derive stellar masses) to a stellar mass limit of $M_\star > 10^8 M_\odot$ is shown in Figure~\ref{fig:results_smd}. We find good agreement with observations~\citep{PerezGonzalez2008,Stark2013,Tomczak2014,Grazian2015,Song2015} taking into account the scatter in observations at $z<2$, and the uncertain but potentially significant contribution of massive quiescent galaxies at low redshift.

\subsection{Luminosity functions and SFR density}
\label{sec:results_LFs}

\begin{figure}[!t]
\centering{
\includegraphics[width=0.49\textwidth]{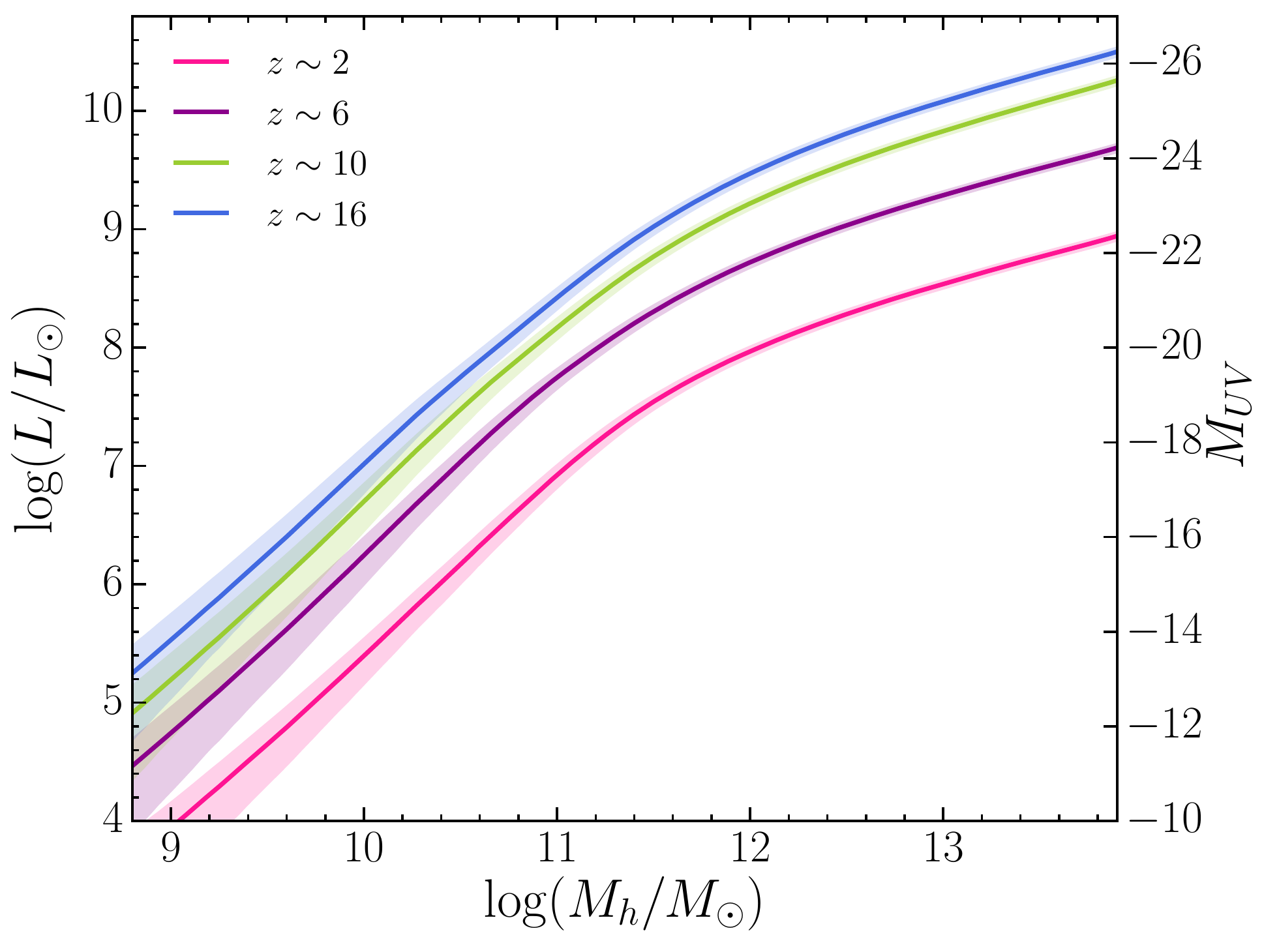}
}
\caption{The relationship between observed galaxy luminosity and halo mass as a function of redshift, $L(M_h, z)$, plotted at $z=2, \, 6, \, 8$, and $16$. Shaded regions show the $1\sigma$ confidence range.}
\label{fig:results_lm}
\vspace{0.2cm}
\end{figure}

The predicted observed $L(M_h,z)$ for a range of redshifts is plotted in Figure~\ref{fig:results_lm}. Our model naturally provides redshift evolution of $L(M_h,z)$ through the evolution of the HMF and halo assembly times. Dust extinction (Section~\ref{sec:model_dust}) significantly affects the high mass end of the relation at low redshift. At $z>10$ we see the high mass end does not evolve much with redshift, motivating the model of~\citet{Mashian2015} which uses an empirical redshift-independent $L(M_h)$. However, there is significant evolution at lower mass, which comprises the greatest contribution to the photon budget available to reionization because of the steep faint-end slope. 

\begin{figure}[!t]
\centering{
\includegraphics[width=0.47\textwidth]{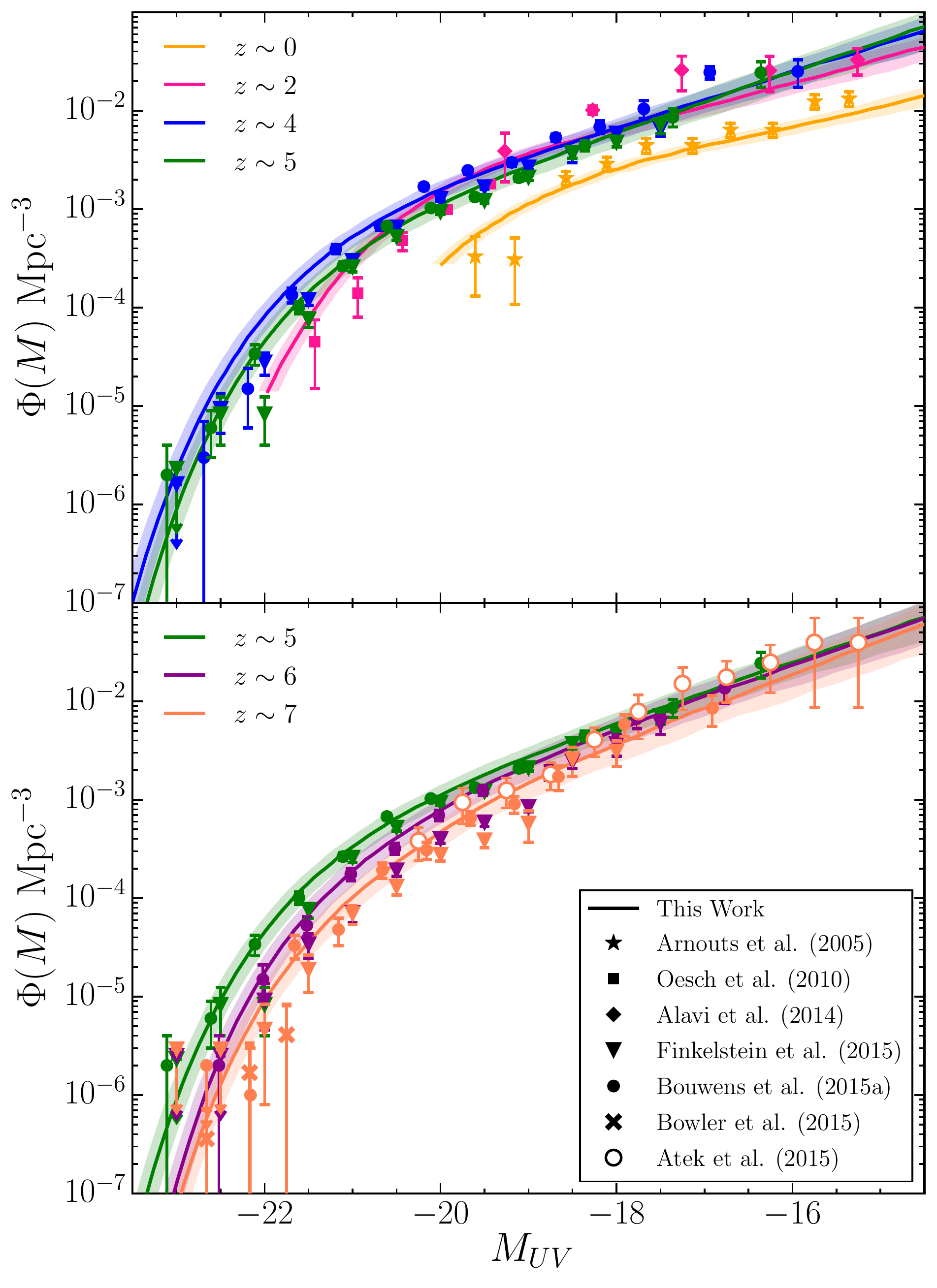}
}
\caption{Predicted UV LFs at low (upper) and intermediate (lower) redshift. We show the LFs using the calibration (see Section~\ref{sec:model_calib}) at $z\sim5$ from~\citet{Bouwens2015a}, with Planck 2015 cosmology~\citep{Planck2015cosmo}, with Planck 2015 cosmology~\citep{Planck2015cosmo}. Points show the binned UV LFs and upper limits from~\citet{Arnouts2005,Alavi2014,Oesch2010,Bouwens2015a,Finkelstein2015,Bowler2015,Atek2015b}. We note that the data from~\citet{Atek2015b} was made public after our model was submitted and illustrates the consistency of our model with observations even at very low luminosity. Shaded regions show the $1\sigma$ confidence range.}
\label{fig:results_LF-lowz}
\vspace{0.3cm}
\end{figure}

\begin{figure}[!t]
\centering{
\includegraphics[width=0.47\textwidth]{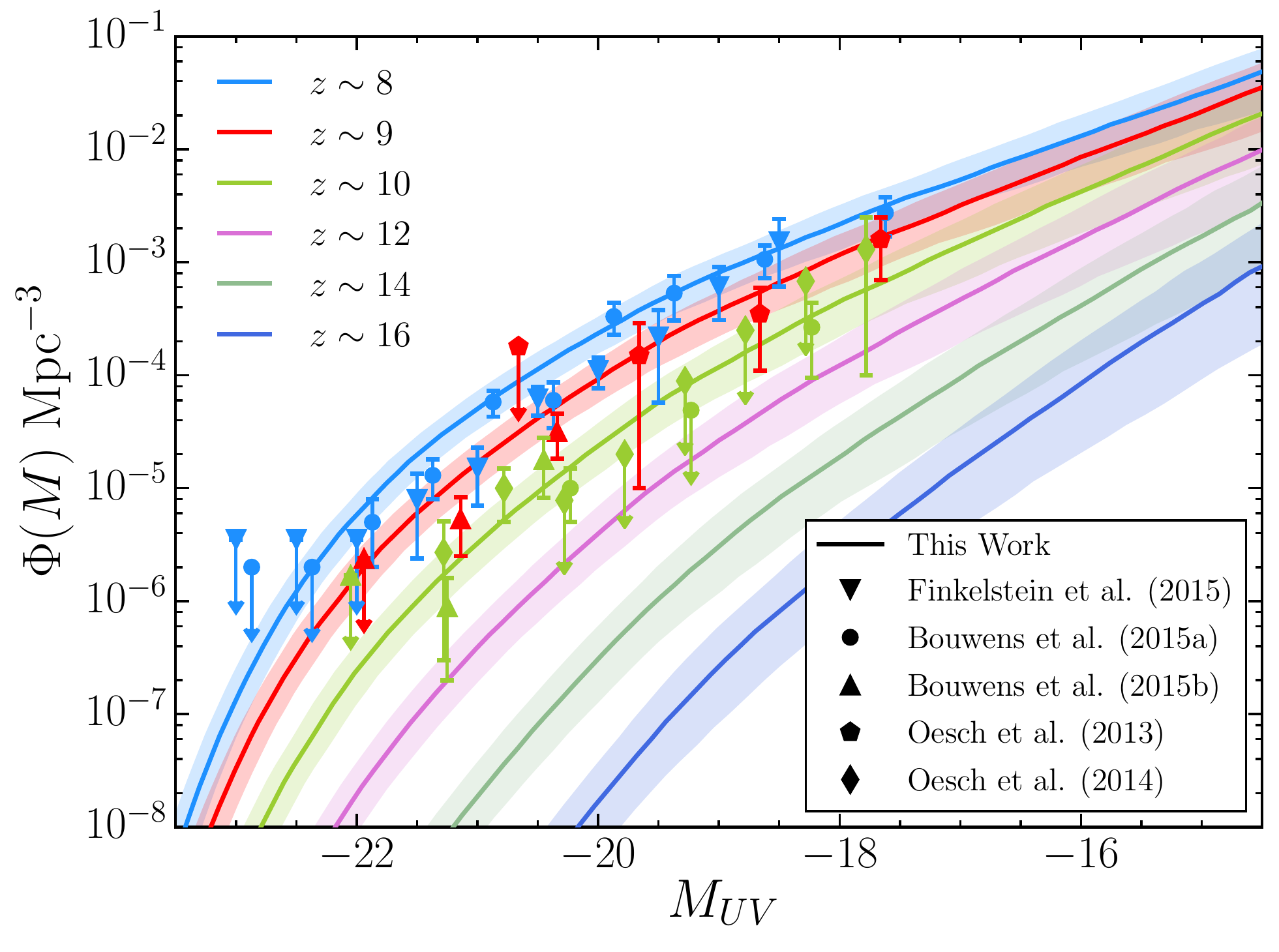}
}
\caption{Predicted UV LFs at high redshift. We show the LFs using the calibration (see Section~\ref{sec:model_calib}) at $z\sim5$ from~\citet{Bouwens2015a}, with Planck 2015 cosmology~\citep{Planck2015cosmo}. Points show the binned UV and upper limits LFs from~\citet{Oesch2013,Oesch2014,Finkelstein2015,Bouwens2015a,Bouwens2015b}. Shaded regions show the $1\sigma$ confidence range.}
\label{fig:results_LF-highz}
\vspace{0.3cm}
\end{figure}

The model UV LFs at $z\leq7$ and $z>7$ are shown in Figures~\ref{fig:results_LF-lowz} and~\ref{fig:results_LF-highz} respectively. The model is remarkably consistent with the observed data ~\citep{Arnouts2005,Oesch2010,Oesch2013,Oesch2014,Alavi2014,Bowler2015,Finkelstein2015,Bouwens2015a,Bouwens2015b}, which is expected due to the success of our previous implementation of this class of models~\citep{Tacchella2013,Trenti2010,Trenti2015}. We find the model marginally overpredicts the bright end of the LF at $z\sim0.3$ and $z\sim2$ by $\sim0.1$ dex due to the difficulty in modeling dust extinction at these redshifts, but the observations are still within $2\sigma$ of our model. In particular, our model predicts a steepening of the faint-end slope at higher redshifts, consistent with the observed trend~\citep{Bouwens2015a}.

At $z>7$, the model describes the most recent observed data ~\citep{Oesch2013,Oesch2014,Finkelstein2015,Bouwens2015a,Bouwens2015b} well, validating our simple approach. Our model predicts the trend of steepening faint-end slope to continue at $z>10$, and number densities to drop rapidly. With \JWST~capabilities, except in an extremely wide-field survey, the UV LF at $z>10$ will be observed as a steep power-law function. This is agreement with the semi-analytic results of~\citet{Behroozi2014} who find a steepening power law slope at faint magnitudes and a significant drop in number densities.

\begin{figure}[!t]
\centering{
\includegraphics[width=0.47\textwidth]{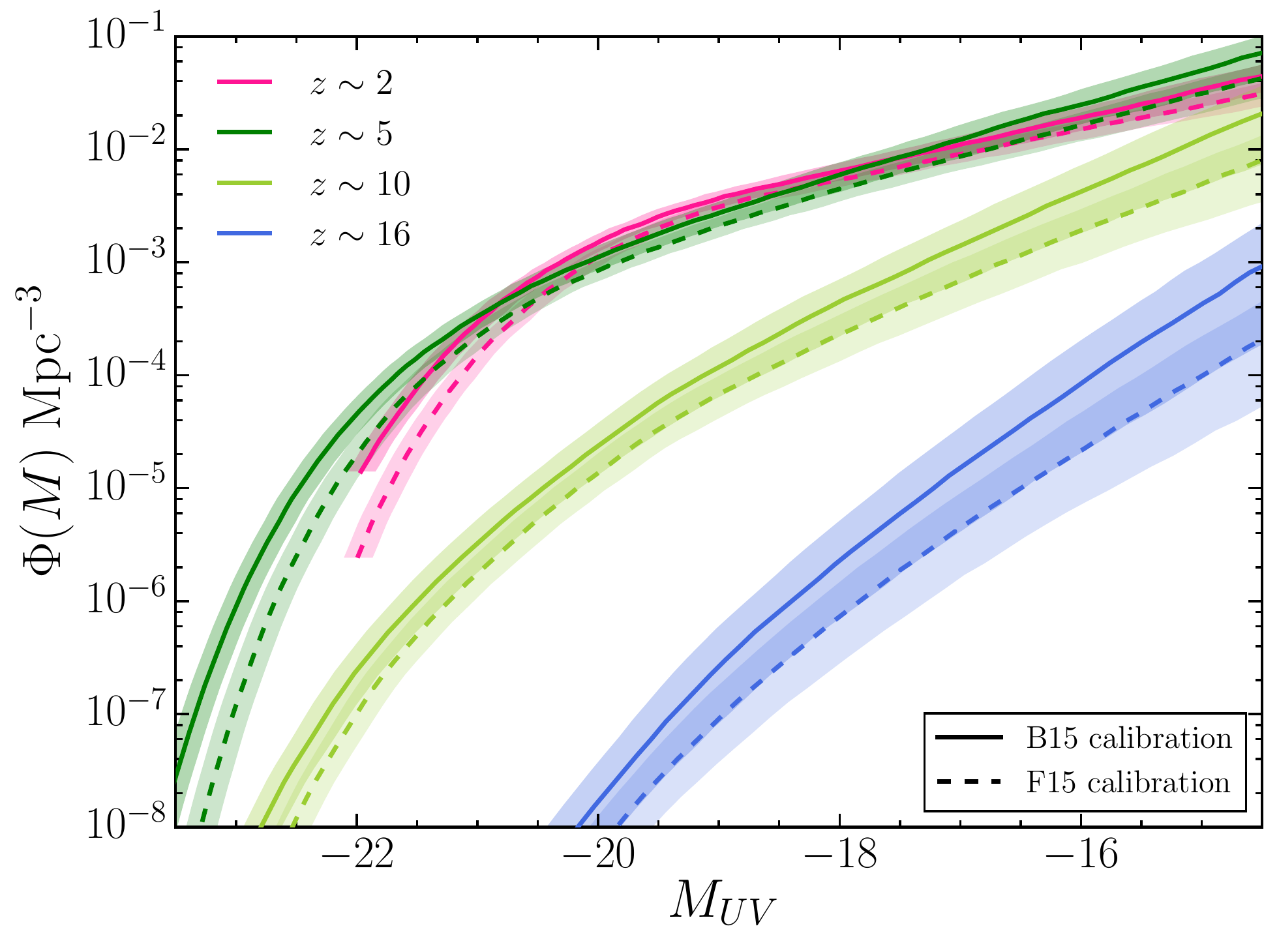}
}
\caption{Predicted LFs at redshifts $z\sim2, \, 5, \, 10, \, 16$ obtained by calibrating (see Section~\ref{sec:model_calib}) our model with the ~\citet{Finkelstein2015} LF at $z\sim5$ (F15, dashed), compared to our reference calibration using the~\citet{Bouwens2015a} LF at $z\sim5$ (B15, solid). Shaded regions show the $1\sigma$ confidence range, highlighting that within the uncertainty of the calibrations, the two approaches yield consistent results.}
\label{fig:Finkelstein_calibration}
\end{figure}

For comparison, we also plot in Figure \ref{fig:Finkelstein_calibration} the model LFs obtained by using the~\citet{Finkelstein2015} $z\sim5$ LF for the calibration. The ~\citet{Finkelstein2015} $z\sim5$ LF has a lower value of $M^*$ than that of~\citet{Bouwens2015a}, so it is not surprising that this calibration produces slightly lower number densities compared to our reference using the ~\citet{Bouwens2015a} LF. However, Figure~\ref{fig:Finkelstein_calibration} clearly shows that  the overall evolution trends are unchanged and that the two calibrations produce LFs which are consistent within one standard deviation both between themselves and the observed data over $0 \simlt z \simlt 10$.

The best-fit~\citet{Schechter1976} function parameters for our LFs are shown in Table~\ref{tab:results_schechter}. The best-fit parameters are in good agreement with observations~\citep{Schmidt:2014p34189,Bowler2015,Oesch2010,Finkelstein2015,Bouwens2015a} given the large degeneracies in Schechter function parameters. Encouragingly, we find the evolution of the derived Schechter parameters is in excellent agreement with the observed evolution~\citep{Bouwens2015a,Bowler2015}: we find $d\alpha/dz \sim -0.1$, $dM^*/dz \sim 0.1$, and $d\log(\Phi^*)/dz \sim -0.3$ between $z\sim4$ and $z\sim8$. We find the evolution of $\alpha$ and $\Phi^*$ between $z\sim8$ and $z\sim16$ to be more dramatic: $d\alpha/dz \sim -0.2$, $dM^*/dz \sim 0.1$, and $d\log(\Phi^*)/dz \sim -0.5$, consistent with the rapid evolution of $\sim10^{10}~M_\odot$ halos in the DM HMF at these redshifts.

\begin{figure}[!t]
\centering{
\includegraphics[width=0.49\textwidth]{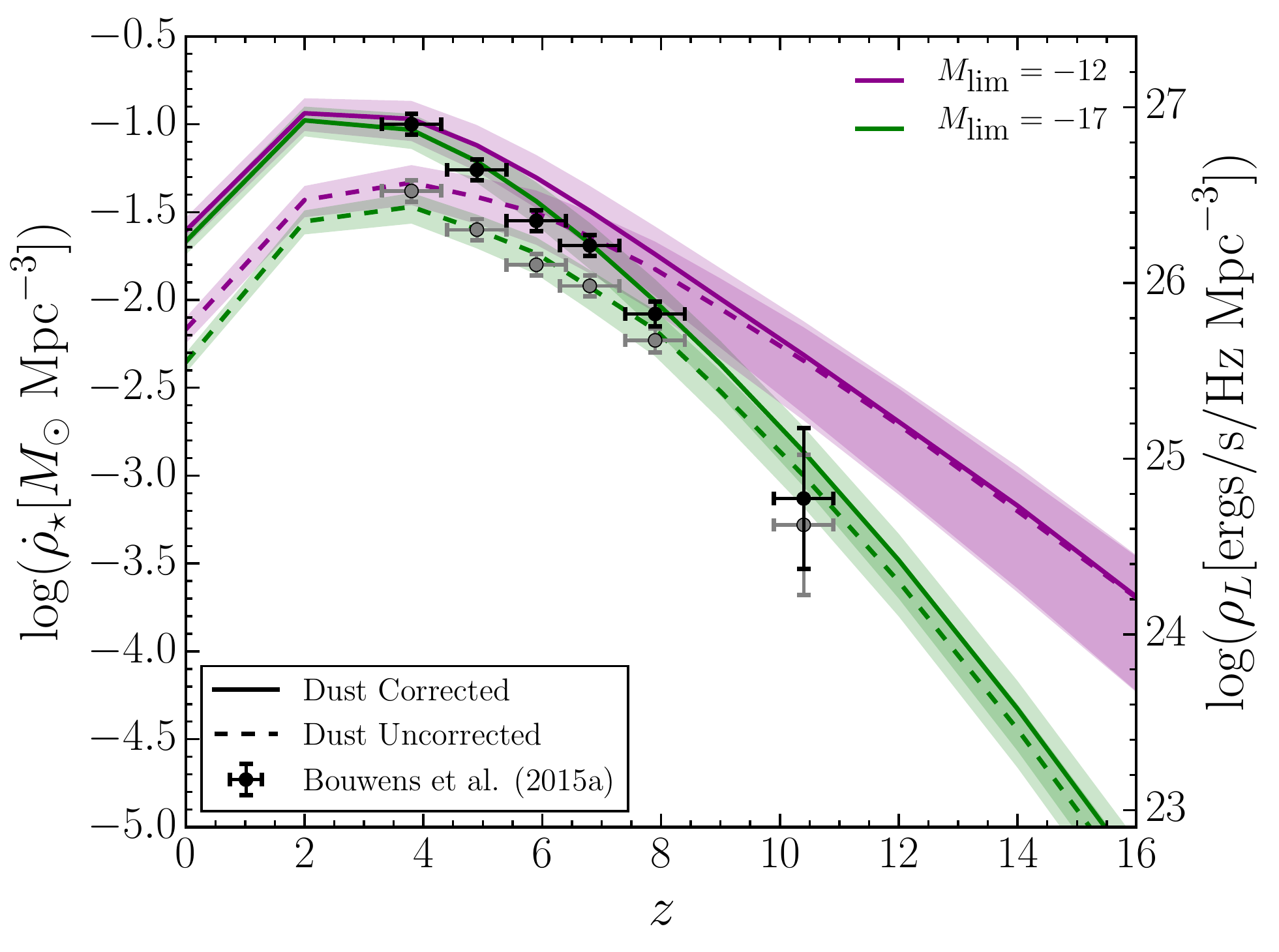}
}
\caption{Luminosity density ($\rho_L$) and cosmic SFR density ($\dot{\rho}_\star$) as functions of redshift, derived by integrating the model UV LFs to magnitude limits of $M_\textsc{ab} = -17$ (green lines) and $M_\textsc{ab} = -12$ (purple lines). The dust corrected SFR densities for the two magnitude limits are shown as solid lines, dust uncorrected SFR densities are shown as dashed lines. The observed SFR densities from~\citet{Bouwens2015a} are shown in black (dust corrected) and grey (dust uncorrected). Shaded regions show the $1\sigma$ confidence range.}
\label{fig:results_sfrd}
\end{figure}

Figure~\ref{fig:results_sfrd} shows the luminosity density and cosmic SFR density as a function of redshift. We calculate the luminosity density by integrating our model LFs down to a magnitude limit. We choose two fiducial limits of $M_\textrm{lim}=-17$ (just fainter than current observational limits) and $M_\textrm{lim}=-12$ (the theoretical mass limit for halos to cool). We calculate the SFR density, $\dot{\rho}_{\star}$ using the empirical relation from ~\citet{Madau1998} where $SFR[M_\odot/\textrm{yr}] = 8.0\times10^{27} L[\textrm{ergs/s/Hz}]$ at 1500 \AA. We plot the densities and observations from ~\citet{Bouwens2015a} both with and without dust correction. At $z\leq8$ the observations are consistent with both magnitude limits, however the observations at $z\sim10$ suggests a significant steepening of the relation at high redshift, as do results from numerical simulations~\citep[][though with large uncertainty]{Genel2014}, which is consistent with our model with $M_\textrm{lim}=-17$. The sample at $z\sim10$ is limited however; more observational data at $z>8$ are needed to confirm this result.

\begin{table}[!t]
\centering{
\caption[ ]{Best-fit Schechter parameters for model LFs}
\label{tab:results_schechter}
\begin{tabular}[c]{cccc}
\hline
\hline
Redshift  	& $\alpha$ 	 		& $M^*$      & $\log(\Phi^* [$mag$^{-1}$Mpc$^{-3}])$   \\
\hline 
$z\sim0$   &   $-1.68\pm0.09$   &   $-19.9\pm0.1$  &   $-2.97_{+0.08}^{-0.07}$  \\
$z\sim2$   &   $-1.46\pm0.09$   &   $-20.3\pm0.1$  &   $-2.52_{+0.09}^{-0.07}$  \\
$z\sim4$   &   $-1.64\pm0.11$   &   $-21.2\pm0.2$  &   $-2.93_{+0.19}^{-0.13}$  \\
$z\sim5$   &   $-1.75\pm0.13$   &   $-21.2\pm0.2$  &   $-3.12_{+0.24}^{-0.15}$  \\
$z\sim6$   &   $-1.83\pm0.15$   &   $-20.9\pm0.2$  &   $-3.19_{+0.25}^{-0.16}$  \\
$z\sim7$   &   $-1.95\pm0.17$   &   $-21.0\pm0.2$  &   $-3.48_{+0.32}^{-0.18}$  \\
$z\sim8$   &   $-2.10\pm0.20$   &   $-21.3\pm0.4$  &   $-4.03_{+0.72}^{-0.26}$  \\
$z\sim9$   &   $-2.26\pm0.22$   &   $-21.2\pm0.4$  &   $-4.50_{+1.36}^{-0.29}$  \\
$z\sim10$   &   $-2.47\pm0.26$   &   $-21.1\pm0.5$  &   $-5.12\pm{0.34}$  \\
$z\sim12$   &   $-2.74\pm0.30$   &   $-21.0\pm0.5$  &   $-5.94\pm{0.38}$  \\
$z\sim14$   &   $-3.11\pm0.38$   &   $-20.9\pm0.5$  &   $-7.05\pm{0.45}$  \\
$z\sim16$   &   $-3.51\pm0.46$   &   $-20.7\pm0.6$  &   $-8.25\pm{0.51}$  \\
\hline
\multicolumn{4}{l}{\textsc{Note.} -- Fit performed between $M_\textsc{ab}=-17.5$ and $M_\textsc{ab}=-22.5$}
\end{tabular}}
\end{table}

\subsection{Forecasts for \JWST~and \WFIRST}
\label{sec:results_forecasts}

We use our model to make forecasts for a representative set of \JWST~NIRCAM high-redshift dropout surveys using the 5 near-IR filters. The surveys (properties summarized in Table~\ref{tab:results_forecasts}) include an ultra-deep (UD) survey of 4 pointings ($\sim40$ arcmin$^2$) exposed in 200 hours per pointing; a medium-deep (MD) survey of 40 pointings exposed in 20 hours per pointing; and a wide-field (WF) survey of 400 pointings exposed in 2 hours per pointing. We assume that the surveys will split the observing time so as to reach equal depth in all five filters, and estimate the limiting magnitude for an $8\sigma$ detection (in a single filter) using the \JWST~Exposure Time Calculator. We also include the effects of gravitational lensing magnification bias from strong lensing in blank fields, which is expected to distort the brightest end of high-redshift LFs~\citep{Mason2015,Wyithe2011}.

\begin{figure}[!t]
\centering{
\includegraphics[width=0.49\textwidth]{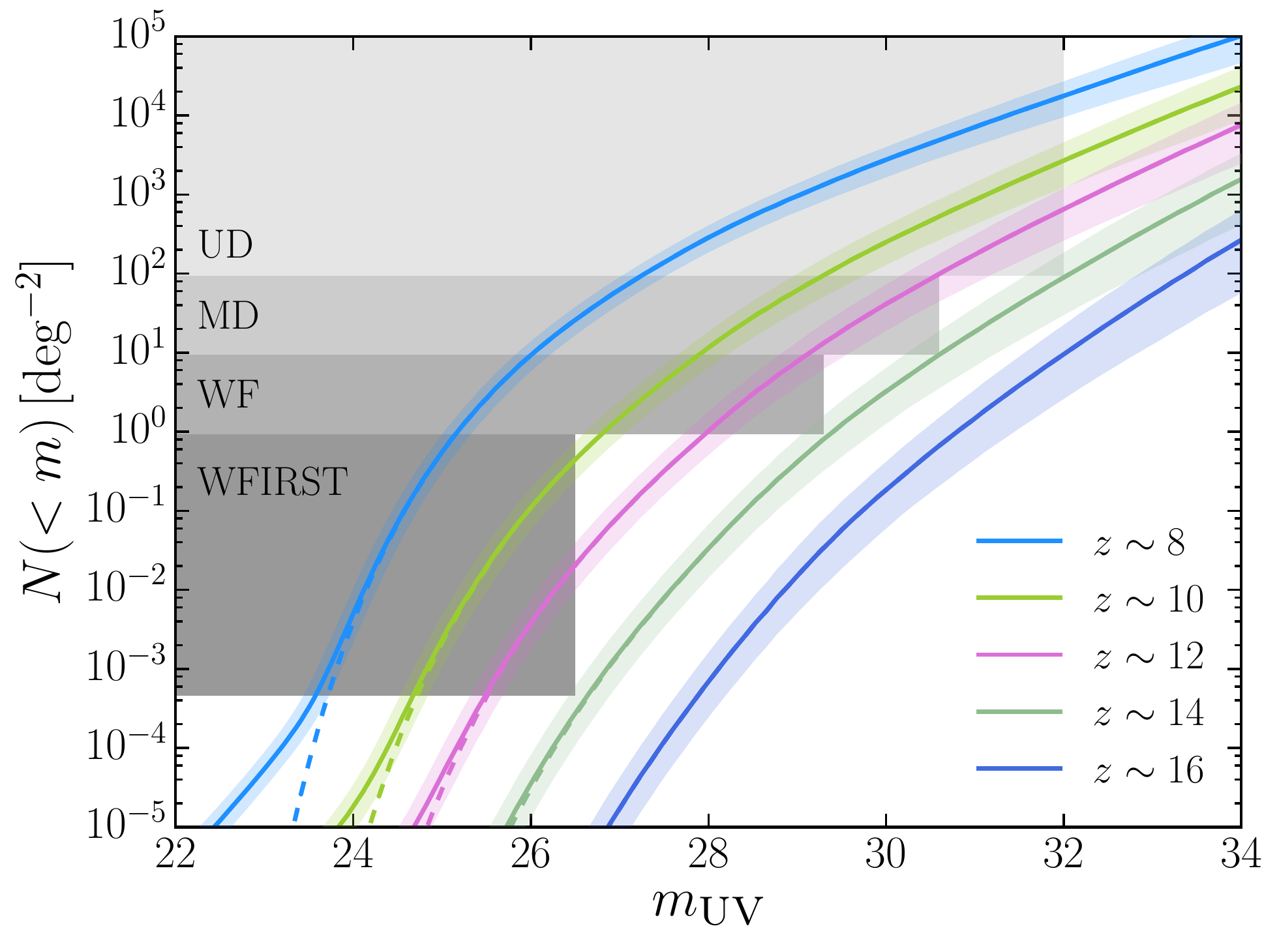}
}
\caption{Predicted number counts of galaxies brighter than apparent magnitude $m_\textsc{UV}$ (rest-frame UV) per square degree for a range of redshifts based on our model LFs. We plot the cumulative number counts including the boost from gravitational lensing magnification bias~\citep{Mason2015,Wyithe2011} as solid lines, and without the magnification bias effect (dashed lines). We plot the estimated coverage of future surveys as shaded regions: 3 mock \JWST~surveys detailed in Section~\ref{sec:results_forecasts} and the \WFIRST~High-Latitude Survey~\citep{Spergel2015}. The calculated number counts are given in Table~\ref{tab:results_forecasts}.}
\label{fig:results_forecast}
\vspace{0.2cm}
\end{figure}

In Figure~\ref{fig:results_forecast} we plot the predicted cumulative number counts for redshifts $8 \leq z \leq 16$ and the regions accessible to these mock \JWST~surveys, as well as the region accessible to \WFIRST~High-Latitude Survey~\citep[HLS,][]{Spergel2015}. The estimated number of dropouts are given in Table~\ref{tab:results_forecasts}.

Our model predicts a significant drop in number density from $z\sim8$ to $z\sim10$ compared to lower redshifts~\citep[which is also seen in the observations,][]{Bouwens2015a,Bouwens2015b}. The drop continues to high redshift, thus we find that no $z\sim16$ galaxies would be detected in our mock \JWST~surveys. To detect 1 galaxy at $z\sim16$ in our UD survey would require $\sim40$ pointings ($\sim400$ arcmin$^2$). 
We find that magnification bias in blank fields does not significantly affect our model even at the brightest observable magnitudes at $z>10$. The magnification bias effect is only noticeable in the exponential part of the LF, which is within reach only at $z~8$, but too weak otherwise for power laws with slope in the range -2 to -3.5 (it is exactly neutral for faint end slope $\alpha = -2$).~\citet{Mason2015} showed the lensing effect was most significant for a Schechter function LF at high redshift (see also \citealt{baronenugent2015}). Thus we expect that without significant strong lensing, i.e. using galaxy clusters as cosmic telescopes~\citep[e.g. the \textit{Hubble Frontier Fields}][]{Yue2014a,Ish+15,CBZ15,Atek2015}, $z>15$ is beyond the reach of \JWST.

\begin{table*}[!t]
\centering{
\caption[ ]{Predicted Number Counts for Example \JWST~and \WFIRST~Surveys}
\label{tab:results_forecasts}
\begin{tabular}[c]{cccccc}
\hline
\hline
Redshift  &  Dropout Filter		& UD $(m_\textrm{lim}=32.0)$ & MD $(m_\textrm{lim}=30.6)$ 	& WF $(m_\textrm{lim}=29.3)$ 	& \WFIRST~$(m_\textrm{lim}=26.5)$ 	\\
		& 					&  $\sim40$ arcmin$^2$ &  $\sim400$ arcmin$^2$ &  $\sim4000$ arcmin$^2$ &  $\sim2000$ deg$^2$ \\
\hline
$z\sim8$  &  F115W   &  $197_{-92}^{+104}$  &  $548_{-225}^{+259}$  &  $1335_{-503}^{+595}$  &  $61370_{-22029}^{+27995}$ \\
$z\sim10$  &  F115W   &  $30_{-17}^{+21}$  &  $52_{-26}^{+33}$  &  $102_{-48}^{+64}$  &  $1026_{-473}^{+701}$ \\
$z\sim12$  &  F150W   &  $6_{-4}^{+5}$  &  $10_{-6}^{+8}$  &  $13_{-7}^{+10}$  &  $47_{-25}^{+41}$ \\
$z\sim14$  &  F150W   &  $0.3_{-0.2}^{+0.4}$  &  $0.4_{-0.2}^{+0.4}$  &  $0.4_{-0.3}^{+0.4}$  &  $0.4_{-0.2}^{+0.4}$ \\
$z\sim16$  &  F200W   &  $0$  &  $0$  &  $0$  &  $0$ \\
\hline
\multicolumn{6}{l}{\textsc{Note.} -- Limiting magnitudes for a $8\sigma$ detection estimated with the \JWST~Exposure Time Calculator and \WFIRST HLS.} \\
\multicolumn{6}{l}{The mock surveys are described in Section~\ref{sec:results_forecasts}. These estimates include the boost from gravitational lensing }\\
\multicolumn{6}{l}{magnification bias in blank fields~\citep{Mason2015,Wyithe2011}.}
\end{tabular}}
\end{table*}

\subsection{Implications for reionization}
\label{sec:results_reion}

The timeline of cosmic reionization depends on the balance between the recombination of free electrons with protons to form neutral hydrogen atoms, and the ionization of hydrogen atoms by Lyman continuum photons emitted by young stars. The UV luminosity density (and therefore, SFR density) at a given redshift allows us to calculate the number of photons available for reionization, and is most sensitive to the faint end of the LF. We can use this to infer the timeline of reionization by calculating the ionized hydrogen fraction, $Q(z)$, as a function of redshift given the following time-dependent differential equation:
\BE  \label{eqn:results_Q}
	\dot{Q} = \frac{\nion}{\nh} - \frac{Q}{\trec}
\EE
where $\nion$ is the comoving number density of ionizing photons, $\nh$ is the comoving number density of hydrogen atoms, and the recombination time of the IGM ~\citep[][and references therein]{Stiavelli2004,Robertson2015} is:
\BE  \label{eqn:results_trec}
	\trec(z) = \left[C\alpha_B(T)n_e(1+z)^3 \right]^{-1}
\EE
where $\alpha_B(T)$ is the case B recombination (i.e. opaque IGM) coefficient for hydrogen, $n_e = (1 + Y_p/4X_p)\nh$ is the comoving number density of electrons (assuming singly ionized He), $X_p$ and $Y_p$ are the primordial hydrogen and helium abundances respectively, and $C=\langle n_H^2 \rangle/\nh^2$ is the ``clumping factor'' which accounts for inhomogeneity in the IGM.

The production rate of ionizing photons can be related to the total UV luminosity density, $\rho_L$ as 
\BE  \label{eqn:results_nion}
	\nion = f_\textrm{esc} \xi_\textrm{ion} \rho_L
\EE
where $f_\textrm{esc}$ is the average fraction of photons which escape galaxies to affect the IGM, and $\xi_\textrm{ion}$ is the rate of ionizing photons per unit UV luminosity, with units Hz/ergs, which depends on the initial mass function, metallicity, age and dust content of the stellar populations. There is an equivalent relation between $\nion$ and SFR density~\citep{Madau1999,Shull2012}, which requires the same stellar population modeling.

All of the parameters involved are difficult to estimate, and may evolve with redshift as reionization progresses and the IGM evolves~\citep{Furlanetto2005,Shull2012}. In this work, we follow~\citet{Schmidt:2014p34189} and use a distribution of parameters. For two limiting magnitudes ($M_\textsc{ab} = -17$, corresponding to currently observable galaxies, and $M_\textsc{ab} = -12$, corresponding to atomic cooling halos) we assume the escape fraction is uniformly distributed between $f_\textrm{esc}=0.1 - 0.3$~\citep{Ouchi2009}, and we use a uniform distribution between $C=1-6$ for the clumping factor. Finally we model $\xi_\textrm{ion}$ as a log-normal distribution with mean $\log{\xi_\textrm{ion}}=25.2$ and standard deviation $0.15$ dex, which was the range obtained by~\citet{Robertson2013} using the standard~\citet{Bruzual2003} models and measurements of the UV spectral slope by~\citet{Dunlop2012}. We assume an IGM temperature of $20,000$~K.

Once the reionization history, $Q(z)$, is known, an important constraint is to compare the electron scattering optical depth with that inferred from CMB observations. The~\citet{Planck2015cosmo} reported a reionization value of $\tau = 0.066\pm0.012$, consistent with instantaneous reionization at $z=8.8_{-1.1}^{+1.2}$. The optical depth as a function redshift is:
\BE  \label{eqn:results_tau}
	\tau(z) = \int_0^z \sigma_T n_e (1+z')^2 Q(z') \frac{c}{H(z')} dz'
\EE
where $c$ is the speed of light, $\sigma_T$ is the Thomson scattering cross section and $H(z)$ is the Hubble parameter.

\begin{figure}[!t]
\centering{
\includegraphics[width=0.47\textwidth]{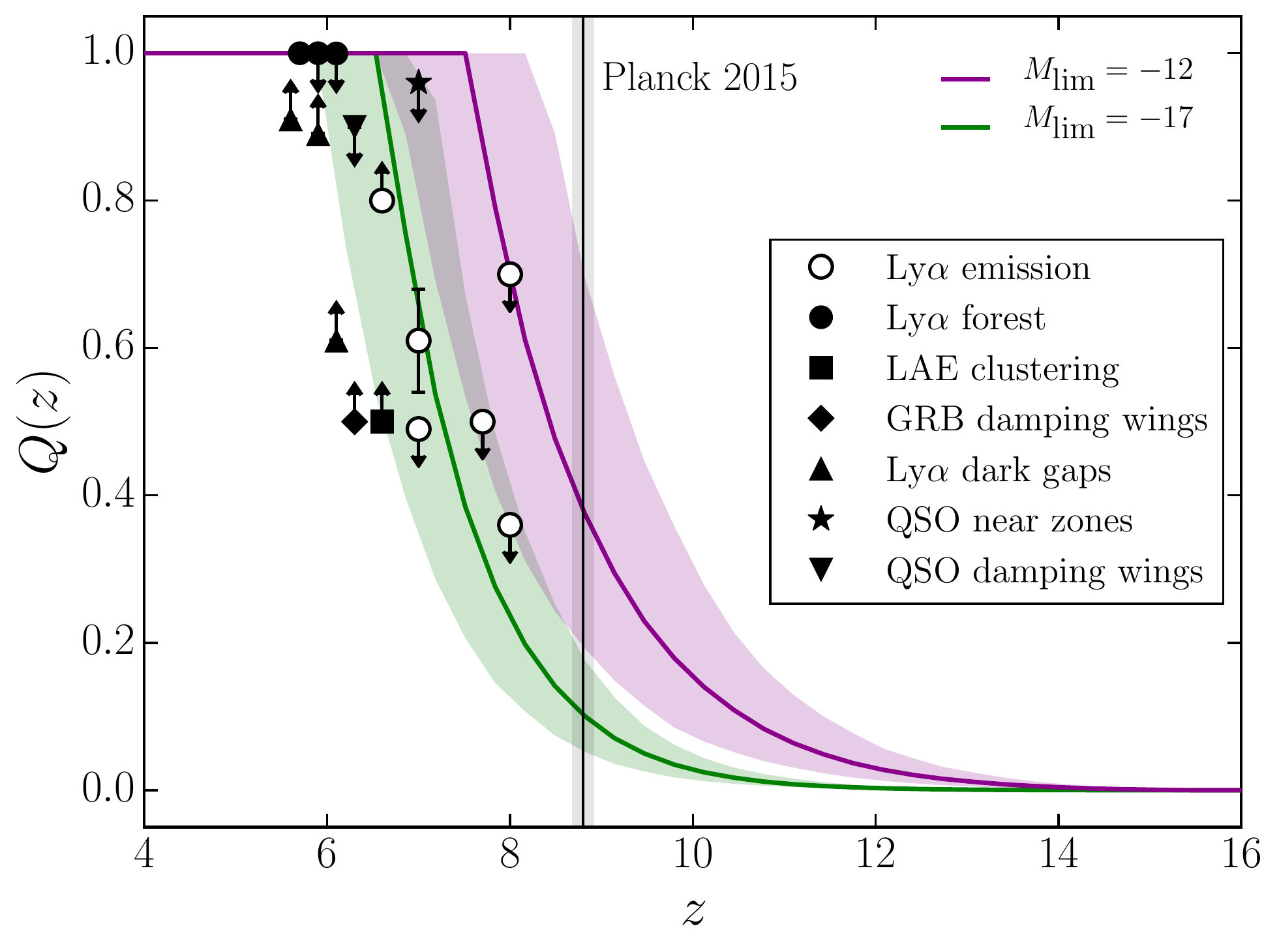}
}
\caption{The fraction of ionized hydrogen as a function of redshift, obtained by solving \Eq{eqn:results_Q} with our model luminosity density. We plot our results from integrating the model UV LFs to two magnitude limits of $M_\textsc{ab} = -17$ (green) and $M_\textsc{ab} = -12$ (purple), with $1\sigma$ confidence regions as shaded regions. We also plot constraints derived from observations of:  \lya~emission from galaxies~\citep[open circles,][]{Ouchi2010,Pentericci2014,Til+14,Fai+14,Sch+14}; the \lya~forest~\citep[filled circles,][]{Fan2006}; the clustering of \lya~emitting galaxies~\citep[square,][]{Ouchi2010}; GRB spectra damping wings~\citep[diamond,][]{McQuinn2008}; dark gaps in the \lya~forest~\citep[upper triangles,][]{McGreer2015}; quasar near zones~\citep[star,][]{Venemans2015}; and quasar spectra damping wings~\citep[lower triangle,][]{Schroeder2013}. We also plot the~\citet{Planck2015cosmo} redshift of instantaneous reionization. We note that the conversion from the \lya~escape fraction to the global ionized hydrogen fraction is uncertain and relies on several model assumptions~\citep{Mesinger2015}.}
\label{fig:results_Q}
\end{figure}

\begin{figure}[!t]
\centering{
\includegraphics[width=0.47\textwidth]{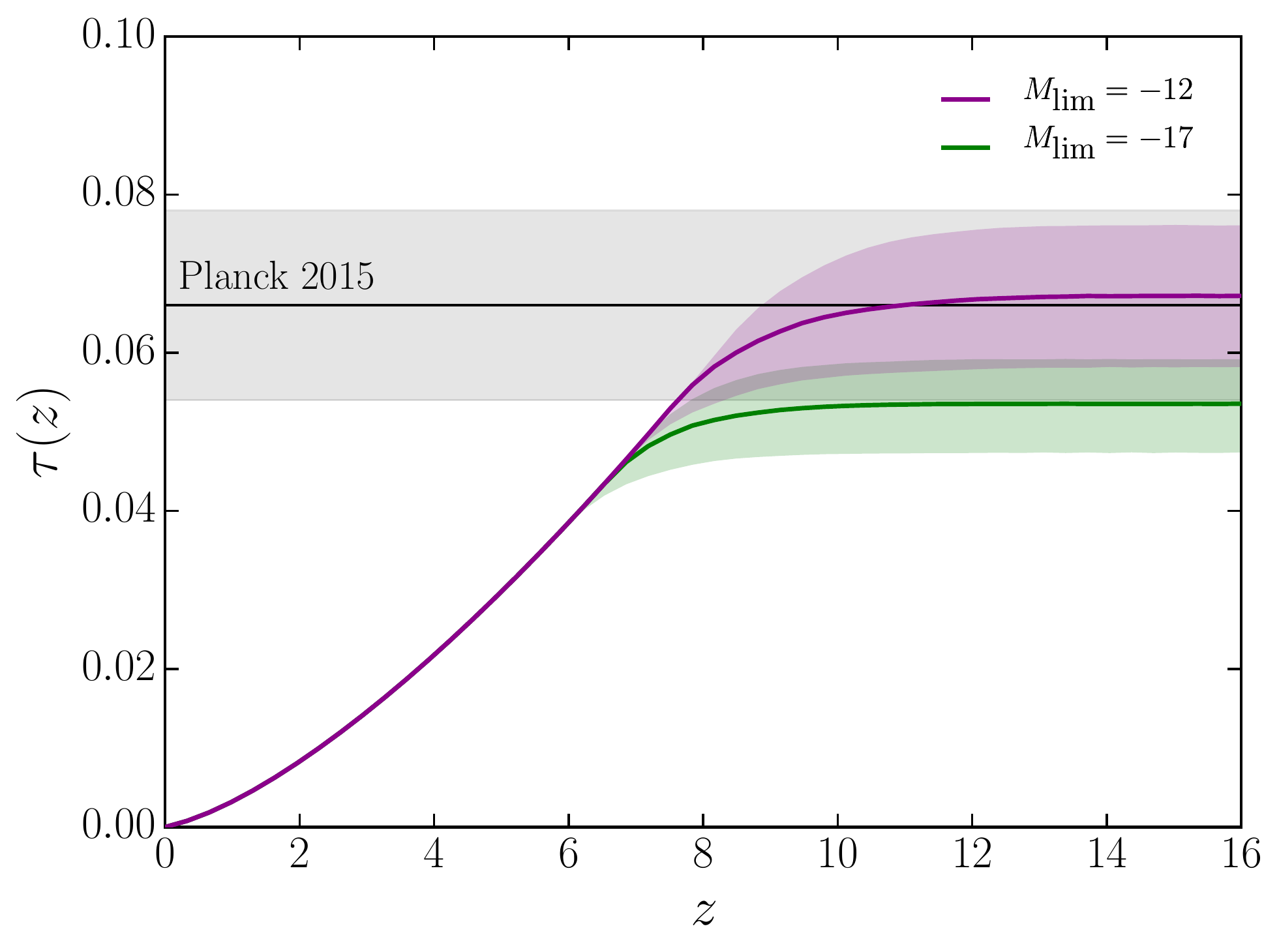}
}
\caption{The electron scattering optical depth, calculated using \Eq{eqn:results_tau} from our derived Q(z). We plot our results from integrating the model UV LFs to two magnitude limits of $M_\textsc{ab} = -17$ (green) and $M_\textsc{ab} = -12$ (purple), with $1\sigma$ confidence regions as shaded regions. We show the reionization optical depth value and its $1\sigma$ confidence levels from~\citet{Planck2015cosmo} in grey.}
\label{fig:results_tau}
\end{figure}

Figure~\ref{fig:results_Q} shows the reionization history: the ionized hydrogen fraction as a function of redshift, obtained by solving \Eq{eqn:results_Q} with our model luminosity density, sampling the distributions of input parameters. Figure~\ref{fig:results_tau} shows the electron scattering optical depth as a function of redshift. For the LF magnitude limit $M_\textsc{ab} = -17$, reionization is complete ($Q=1$) by $z_\textrm{reion}=6.86^{+0.32}_{-0.66}$, with $\tau(z_\textrm{reion}) = 0.042^{+0.008}_{-0.002}$. For the LF magnitude limit $M_\textsc{ab} = -12$, reionization is complete by $z_\textrm{reion}=7.84^{+0.65}_{-0.98}$, with $\tau(z_\textrm{reion}) = 0.056^{+0.007}_{-0.010}$. 

The fainter magnitude limit, corresponding to atomic cooling halos of mass $M_h \sim 10^9 M_\odot$, is fully consistent with the Planck results, considering the uncertainty in the reionization model parameters. This calculation shows that ultrafaint galaxies can in principle provide enough photons to fully reionize the universe by $z\sim6$ to match observations of the \lya~forest~\citep{Fan2006}. Both magnitude limits are broadly consistent with a range of constraints from observations, within the reionization model uncertainty: UV luminosity densities~\citep{Finkelstein2012} for observable galaxies; quasar near zones~\citep{Venemans2015}; quasar spectra damping wings~\citep{Schroeder2013}; GRB spectra damping wings~\citep{McQuinn2008}; transmission ~\citep{Fan2006} and dark gaps~\citep{McGreer2015} in the \lya~forest; and the clustering of \lya~emitting galaxies~\citep{Ouchi2010}.

Qualitatively, the non-negligible neutral fraction predicted by our model at $z\simgt7$ is consistent with the observed high optical depth of \lya~\citep[][K. B. Schmidt et al. 2015, ApJ submitted]{Ouchi2010,Tre+13,Pentericci2014,Sch+14,Til+14,Fai+14}, however the conversion from the \lya~emission fraction to the volume filling factor of ionized hydrogen is difficult and requires several assumptions \citep{Mesinger2015}. In particular, to make constraints on reionization it is generally assumed that there are no changes in galaxy and the \lya~emission line properties, which necessitates a rapid evolution of the global ionization fraction between $z\sim6$ and $z\sim7$. However, recent studies have shown that the rapid decline in the \lya~escape fraction at these redshifts cannot result only from the changing IGM attenuation~\citep{Mesinger2015} but could also be explained by the co-evolution of the escape fraction of ionizing photons, $f_\textrm{esc}$,~\citep{Dijkstra2014}. Thus, the uncertainties in the ionization fraction from the \lya~optical depth shown in our plot are likely underestimated, since they do not include these systematic effects.

\section{Summary and Conclusions}
\label{sec:conc}

We have presented a simple model for the evolution of the UV LF from $0\simlt z \simlt 16$, assuming that the average star formation history of galaxies is set by their halo mass and by the redshift (through the halo assembly time), so that halos of the same mass have the same stellar mass content independent of redshift. Our model builds upon previous similar implementations, but here we extended our framework to construct a self-consistent model which is capable of following the evolution of the star formation even when the halo assembly times become very short (at $z\simgt10$). 

Our key findings are as follow: 

\begin{enumerate}

\item Our model UV luminosity functions are very successful in matching observations at all redshifts where data are available ($0\simlt z \simlt 10$). Overall, we find that the shape of the LF is well described by a Schechter function with faint-end slope increasing with redshift. This trend continues at higher redshift, and we use the model to make predictions for LFs at $z>10$, finding a faint-end slope $\alpha\sim-3.5$ at $z=16$.

 \item Our model reproduces the observed cosmic SFR density well, indicating a sharp decline at $z>8$ with a magnitude limit of $M_\textsc{ab}=-17$, consistent with observed data at $z\sim10$.

\item Compared to previous more basic models~\citep{Tacchella2013,Trenti2015}, we find that the self-consistent inclusion of earlier periods of star formation does not significantly affect the total UV luminosity at a given halo mass and redshift, but it allows us to better reproduce the observed average stellar ages and stellar mass density of high redshift galaxies.

\item Taking advantage of the ability of the model to make predictions at the earliest times, we investigate the expected galaxy detections for future ultra-deep, medium-deep and wide-field surveys with \JWST, and \WFIRST. We predict that $z\sim14$ galaxies over a range of luminosities are in reach of these surveys. However, significant strong lensing magnification will be needed to push beyond $z>15$.

\item Finally, we investigate the implications of our model for the reionization process and find that reionization is complete by $z_\textrm{reion}=7.84^{+0.65}_{-0.98}$, under the assumption that the LF extends down to a minimum galaxy luminosity of $M_\textsc{ab}=-12$ ($M_h\sim10^9M_\odot$), with $\tau(z_\textrm{reion}) = 0.056^{+0.007}_{-0.010}$. Overall our model is consistent with the~\citet{Planck2015cosmo} results and with ultrafaint galaxies being the dominant sources of reionization, despite the fact that this population is currently not detected via direct imaging (but inferred indirectly through GRB host galaxy searches at $z>6$).

\end{enumerate}

\acknowledgments

This work was supported by the \HST~BoRG grants GO-12572, 12905, and 13767, and the \HST~GLASS grant GO-13459. T.T. acknowledges support by the Packard Foundation through a Packard Fellowship.

\bibliographystyle{apj}
\bibliography{bibtexlibrary}

\end{document}